\documentstyle[11pt,newpasp,twoside,epsf]{article}
\def\edcomment#1{\iffalse\marginpar{\raggedright\sl#1\/}\else\relax\fi}
\reversemarginpar

\newcommand{\ie}{\emph{i.e.}\ }
\newcommand{\etal}{\emph{et al.}\ }
\newcommand{\eg}{\emph{e.g.}\ }
\newcommand{\cf}{\emph{c.f.}\ }

\newcommand{\msun}{M$_\odot$}
\newcommand{\kms}{km~s$^{-1}$}
\newcommand{\lsim}{{\, \lower2truept\hbox{
${< \atop\hbox{\raise4truept\hbox{$\sim$}}}$}\,}}
\newcommand{\gsim}{{\, \lower2truept\hbox{
${> \atop\hbox{\raise4truept\hbox{$\sim$}}}$}\,}}

\begin{document}
\title{Primordial Stellar Populations}
\author{Nino Panagia}
\affil{ESA/Space Telescope Science Institute, 3700 San Martin Drive,
Baltimore, MD 21218}

\begin{abstract}
We review the expected properties of the first stellar generations in
the Universe. In particular, we consider and discuss the diagnostics,
based on the emission from associated HII regions, that permit one to
discern {\it bona fide} primeval stellar generations from the ones
formed after pollution by supernova explosions has occurred. We argue
that a proper characterization of truly primeval stellar generations
has to be based on spectra that show simultaneously (a) the presence of
high intensities and equivalent widths of  Hydrogen and Helium emission
lines , such as Ly-$\alpha$ and HeII 1640\AA, and (b) the absence of
collisionally excited metal lines, mostly from the first heavy elements
to be synthetized in large quantities, i.e. C and O. These atomic
species are expected to produce emission lines, such as CIII] 1909\AA,
OIII] 1666\AA, [OIII] 5007\AA, {\it etc.}, with intensities above 10\%
the intensity of H$\beta$ already for metallicities as low as
0.001Z$_\odot$.  The expected performance of the NASA/ESA/CSA NGST for
the study and the characterization of primordial sources is also
discussed.

\end{abstract}

\section{Introduction}

The first generation of stars in the Universe marks the beginning of
the evolution that eventually lead to the world as we know it.  Before
the formation of {\it any} star, the chemical composition is the one
produced by the Big Bang, i.e. mostly Hydrogen (about 93\% by number)
and Helium (about 7\% by number) and traces of other light elements,
such as Deuterium, Lithium, etc. (\eg Truran, this Conference).  The
stars of the first  generation, the so-called {\it ``population III"}
stars, will reflect the lack of metals in having considerably higher
effective temperatures than stars of equal mass but with appreciable
quantities of heavy metals (see \eg Castellani, Chieffi \& Tornamb\`e
1983, and many more; see Marigo \etal~2001 for extensive calculations
and a  thorough review).  On this basis, Tumlinson, Giroux \& Shull
(2001) and Bromm, Kudritzki \& Loeb (2001) have argued that the high
temperatures of the most massive stars will produce HII regions in
which part of the helium is twice ionized.  From this, they predict
intensities of the HeII 1640\AA\/ line so uniquely high
(I(1640\AA)/I(H$\beta$)$>0.5$)  that the detection of a strong  HeII
1640\AA\/ line in the spectrum of a high redshift galaxy should
constitute clear evidence for a primeval stellar generation.  These
predictions are heavily based on the assumption that primordial stars
are {\it all} massive stars or, in other words, that a ``top-heavy"
Initial Mass Function (IMF) is  appropriate for population III stars.
Undoubtedly, it is intuitive that the absence of metals  makes the
temperature in primordial star forming clouds higher than it is in the
local Universe, so that the Jeans mass is higher and the formation of
high mass stars should be highly favored while the formation of low
mass stars is expected to be greatly reduced or completely inhibited.
On the other hand, the discovery of Milky Way halo stars with
metallicities as low as $[Fe/H]\sim-4$ (\eg Depagne \etal 2002, and
references therein) casts disturbing doubts about the general validity
of a top-heavy IMF in the early Universe.

   \begin{figure}
   \plotfiddle{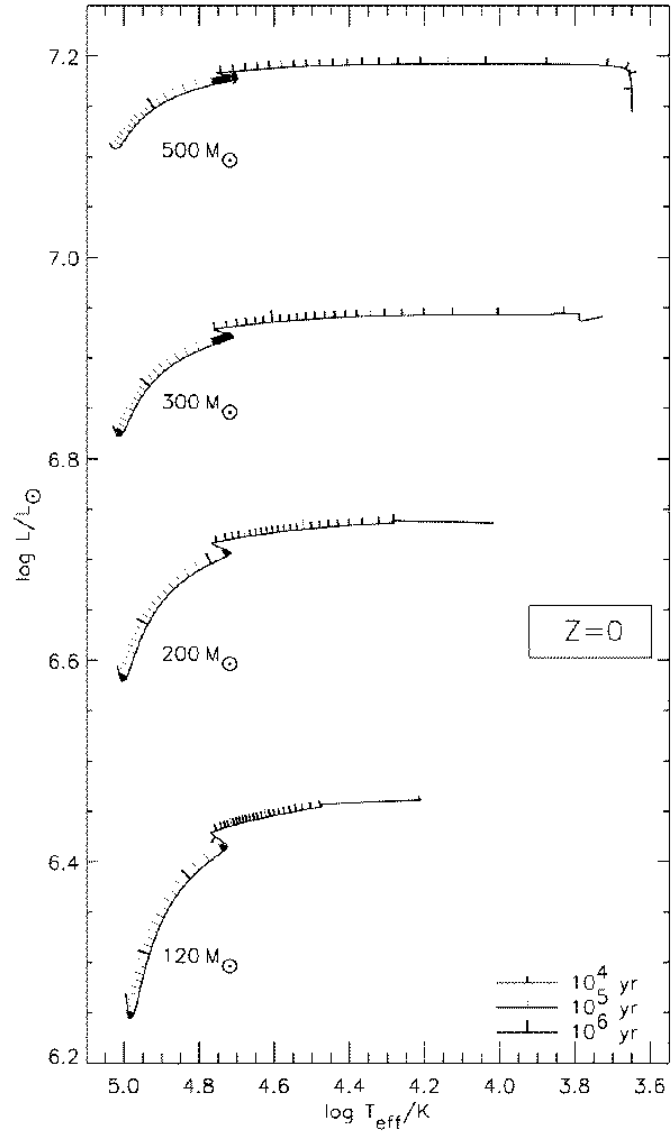}{14cm}{0}{60}{60}{-180}{-20}
   \caption{The HR diagram for zero-metallicity massive stars (adapted
   from Baraffe \etal~2000). }
   \end{figure} 

Another possible problem is that the phase of ``primordial" abundance
may be so short that the probability of catching any proto-galaxy in
that phase could be exceedingly low.  In fact, in a short time after
the very beginning of star formation (about 3 million years after the
birth of the first massive stars in the Universe)  supernova explosions
will start polluting the environment and the conditions will suddenly
change.

It is  then clear that identifying and characterizing the properties of
the {\it second} generation of stars, \ie stars formed in the metal
enriched environment just after the first episode of SN pollution,  is
at least as important as identifying the truly primeval stellar
population.

In the following section we will show that a proper characterization 
of truly primeval stellar generations has to be based on the
simultaneous presence of high intensities and equivalent widths of
suitable lines of Hydrogen and Helium, such as Ly-$\alpha$ and HeII
1640\AA~(and/or their rest-frame optical counterparts, H$\alpha$ and
HeII 4686\AA), {\it and} the absence of collisionally excited metal
lines, mostly from the first heavy elements to be synthetized in large
quantities, \ie C and O, which are expected to have lines such
CIII]~1909\AA, OIII]~1666\AA, [OIII]~5007\AA, {it etc.,} with
intensities above 10\% the intensity of the H$\beta$ line already for
metallicities as low as $\sim10^{-3}$ solar.

\section{Primordial Stars: Expected Properties}

As mentioned before, the current paradigm is that at zero metallicity
the Jeans mass in star forming clouds is much higher than it is in the
local Universe, and, therefore, the formation of massive stars, say,
100 \msun\/ or higher, is highly favored (\cf Figure 1). The spectral
distributions (SED) of these massive stars are characterized by
effective temperatures on the Main Sequence (MS) around $10^5$~K
(Tumlinson \& Shull 2000, Bromm \etal~2001, Marigo \etal~2001).  Due to
their temperatures these stars are very effective in ionizing hydrogen
and helium. It should be noted that zero-metallicity stars of all
masses are expected to have essentially the same MS luminosities as,
but to  be much hotter than their solar metallicity analogues
(Tumlinson \& Shull 2000, Bromm \etal~2001, Marigo \etal~2001; see \eg
Fig. 2).  Note also that only stars hotter than about 90,000~K are
capable of ionizing He twice in appreciable quantities, say, more than
about 10\% of the total He content (\eg Oliva \& Panagia 1983,
Tumlinson \& Shull 2000).  As a consequence even the most massive
population III stars can produce HeII lines only for a relatively small
fraction of their lifetimes, say, about 1~Myr or about 1/3 of their
lifetimes (Baraffe, Heger \& Woosley 2001;  \cf Fig. 1). 

   \begin{figure}
  \plotfiddle{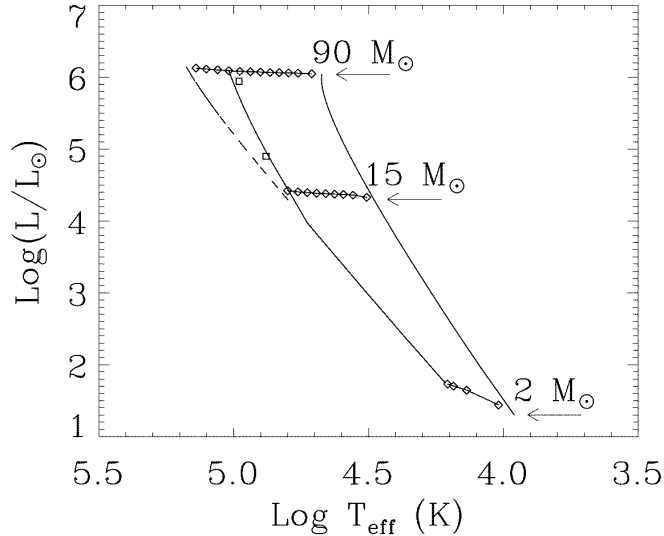}{7.cm}{-90}{38}{38}{-150}{220}
   \caption{The HR diagram for stars of metallicities from solar value
   (right-most curve) to zero (left-most curve).  The diamonds indicate
   the location of 2, 15 and 90 \msun\/ models for metallicities
   decreasing by subsequent factors of ten (adapted
   from Tumlinson \& Shull~2000). }
   \end{figure} 

The second generation of stars forming out of pre-enriched material
will probably have different properties because cooling by metal lines
may become a viable mechanism and  stars of lower masses may be
produced (Bromm \etal 2001). On the other hand, if the metallicity is
lower than about $5\times 10^{-4}$Z$_\odot$,  build up of H$_2$ due to
self-shielding may occur, thus  continuing the formation of very
massive stars (Oh \& Haiman 2002).

Thus, it appears that in the zero-metallicity case one may  expect a
very top-heavy Initial Mass Function (IMF). It is less clear whether
the second generation of stars is also top-heavy or characterized by a
more normal IMF. We will discuss both possibilities.

Population III objects are characterized by nucleosynthetic patterns
different from those produced by ordinary stars (Heger \& Woosley 2002,
Oh \etal 2001).  In principle, such patterns could be identified by
studying low metallicity objects at low redshifts.

\section{Primordial HII Regions}

   \begin{figure}
   \plotone{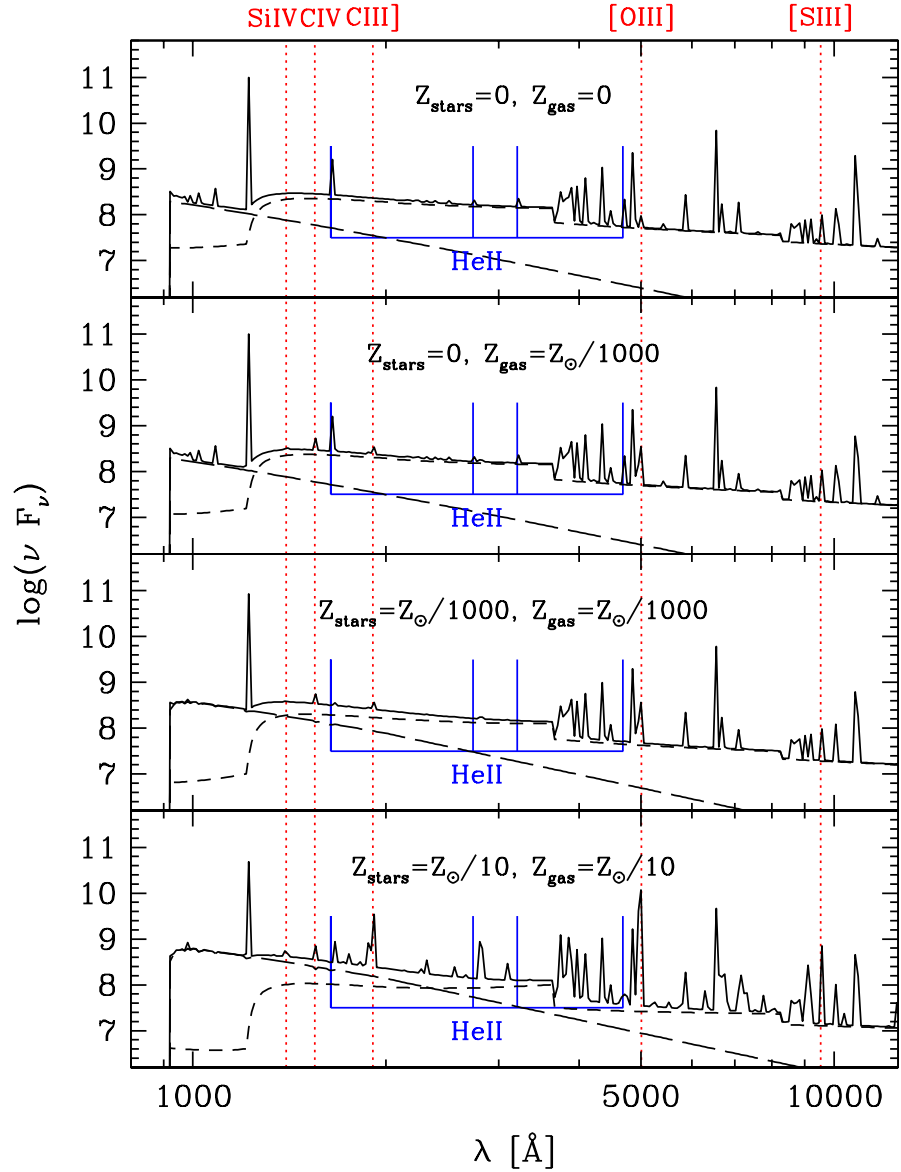}
   \caption{The synthetic spectrum of a zero-metallicity HII region
(top panel) is compared to that of HII regions with  various
combinations of stellar and nebular metallicities (lower panels). The
long dashed lines represent the stellar continuum while the short
dashed lines represent the nebular continuum. Note how the latter
dominates the continuum in the zero-metallicity case for
$\lambda>1216$\AA. }
   \end{figure} 

Given the  high effective temperatures of zero-metallicity stars we can
expect high ionizing photon fluxes for both hydrogen and helium. By the
same token, we also expect low optical  and UV fluxes.  This is because
the optical/UV domains fall  in the Rayleigh-Jeans tail  of the
spectrum where the flux is proportional to the first power of the 
effective temperature, $T_{eff}$,  so that, for equal bolometric
luminosity, the actual flux scales like $T_{eff}^{-3}$.  Therefore, an
average increase of effective temperature of a factor of $\sim2$ will
give a reduction of the optical/UV flux by a factor of $\sim8$. As a
result,  one should expect the rest-frame optical/UV spectrum of a
primordial HII regions to be largely dominated by its nebular emission
(both continuum and lines).

More quantitatively, we have computed an extensive grid of models to
study the detailed properties of primordial HII regions (Panagia \etal
2002). The calculations were made using  Cloudy90 (Ferland \etal
1998),and exploring a wide range of densities (1--10$^6$~cm$^{-3}$),
metallicities (10$^{-6}$--3~Z$_\odot$), and effective temperatures
(30,000--150,000~K).  For the SED of the ionizing stars we considered
both model atmospheres, as provided by Cloudy90 database and black body
distributions. We have also considered the cases of both groups of
identical stars and clusters of MS stars with a Salpeter IMF. Although
a variety of relative elemental abundances were investigated, in most
models we varied the metallicity by simply scaling all metal abundances
by one and the same factor, relative to the solar abundances as listed
in the Cloudy90 manual.   The main results of our model calculations
for the case of zero metallicity are:
\begin{itemize}
\item  The electron temperature is much higher than in galactic HII
regions, in excess of 20,000~K in all cases.  This is because both the
average energy of the ionizing photons is considerably higher (by about
3-4 eV), and most of the cooling is provided by collisional excitation 
of hydrogen levels, whose {\it lowest} excitation potential is about
10.2~eV.
\item As a corollary the intensities of all hydrogen lines are enhanced
relative to  the classical Baker \& Menzel (1938) case B recombination
regime.  In particular, thanks to collisional excitation, the 
Ly-$\alpha$ line is about twice as predicted from recombinations alone,
and its luminosity may amount to as much as 46\% of the total
luminosity of the ionizing star(s).  Similarly, the H$\alpha$ to
H$\beta$ intensity ratio goes to about 3.2, \ie about 10\% higher than
the standard case B ratio.
\item As anticipated, the continuous emission is largely dominated  by
the nebular continuum at all wavelengths longer than Ly-$\alpha$.  As a
consequence, the resulting SED is much flatter than the stellar one,
and is also appreciably  flatter than that of solar abundance HII
regions (\cf Fig. 3).  This effect  results both from the higher
electron temperature that makes the $f-b$ and $f-f$ spectra
considerably flatter, and from the collisional excitation of the $2s$
level of hydrogen, whose decay produces the well known two-photon
emission that extends from 0 energy up to E(Ly-$\alpha$) and dominates
the nebular emission between 1216\AA\/ and, say, 5000\AA, at least.
\item Helium is mostly once ionized but, as long as the effective 
temperature of the ionizing stars is above 90,000~K, emission lines of
HeII can also be detected.  In particular, for $T_{eff}=$100,000~K, the
fraction of twice ionized He is 29\% for a model atmosphere SED and
10\% for a black body SED. Correspondingly, the intensities of the HeII
lines  at 1640\AA\/ and 4686\AA\/ are about 1.26 and 0.16 (for model
atmospheres), and  0.62 and 0.078 (for black bodies)  the intensity of
the H$\beta$ line, respectively.
\item There are no metal lines, of course.
\item There is no dust absorption of any kind, {\it but} both the
higher H$\alpha$ to H$\beta$ intensity ratios and the flatter
rest-frame UV continuum mimick the effects of dust extinction at a
level equivalent to E(B-V)$\simeq$0.1 if a Milky Way average extinction
law is adopted. This, in turn, may (incorrectly) be taken as evidence
for a substantial metallicity, up to 0.1\/Z$_\odot$, in these
primordial regions.
\end{itemize}

   \begin{figure}
   \plotfiddle{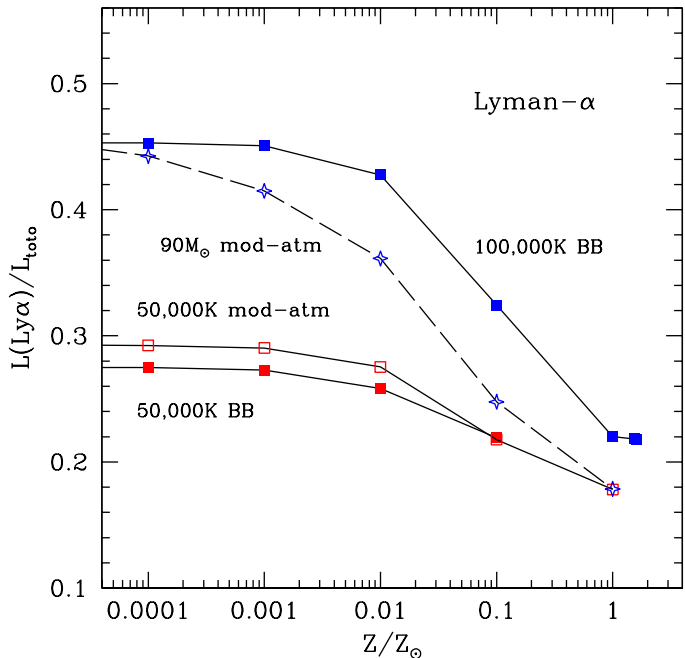}{7.99cm}{0}{50}{50}{-150}{-100}   
   \caption{ Ly-$\alpha$ luminosity in units of the total stellar
   luminosity as a function of metallicity for HII regions ionized by 
   stars with 50,000 and 100,000~K black-body SED (filled squares), 
   50,000~K model-atmosphere SED (open squares), and a  90 \msun\/star 
   with model-atmosphere SED and effective temperatures as computed by
   Tumlison \& Shull (2000).
}
   \end{figure} 
   \begin{figure}
   \plotfiddle{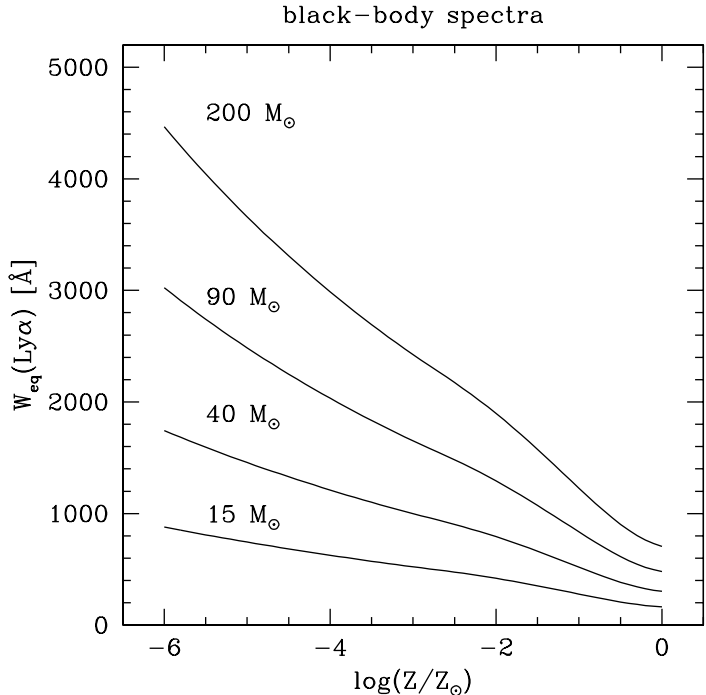}{7.99cm}{0}{50}{50}{-150}{-100}   
   \caption{ Ly-$\alpha$ equivalent widths for HII regions ionized by
stars with a range of masses and metallicities. The results obtained
for black bodies or stellar atmospheres are very similar.}
   \end{figure} 

\section{Low Metallicity HII Regions}

As soon as the first massive star reaches the end of its lifetime
(about 3 Myrs; \cf Baraffe \etal 2001, Marigo \etal 2001) and  explodes
as a core collapse (type II)  supernova (SNII), its environment is
polluted and may show the effects of the ensuing metal enrichment.  It
is important to realize that a few SN explosions are more than enough
to pollute a big cloud appreciably.  For instance, if we assume that,
similarly to local Universe SNII, a primordial SNII explosion ejects
about 10 \msun\/ of metals with a total kinetic energy of few
10$^{51}$ergs, it is easy to verify that the ejecta will be stopped
(actually, thermalized) in about 1 Myr after sweeping, and mixing with 
about 10$^6$ \msun\/ of gas of the parent cloud.  This corresponds to
an average metallicity of $Z\simeq 0.00001 \simeq Z_\odot$/2000. 
Therefore, we can expect that eventually, when all massive stars of the
first generation have gone SNII, even a cloud with as much as 10$^7$
\msun\/ in gas will be polluted to a level of about
10$^{-3}$Z$_\odot$.  Interestingly enough, this is the same level of
metal enrichment as expected for the Universe at the epoch of
re-ionization (\eg Miralda-Escud\'e \& Rees 1998).  While a little
confusing (which is which? which is first?) this fact simplifies the
job of recognizing ``pristine" stars and clouds: whatever is above the
limit of   $\sim$10$^{-3}$Z$_\odot$ is the product of one or more
episodes of metal enrichment.  

The properties of HII regions with various amounts of metal enrichment
can be inferred from Figures 3-6. In particular,  Figure 3, illustrates
four different scenarios (top to bottom): (1) both stars and gas are
metal free, (2) the stars are primordial but the gas is already
enriched to  10$^{-3}$Z$_\odot$, (3) both stars and gas have a
metallicity of  10$^{-3}$Z$_\odot$, and (4) the metallicity of both
stars and gas is 0.1Z$_\odot$, similar to the SMC metallicity.  We see
that, as long as primordial massive stars dominate the ionization ((1)
and (2)), HeII emission lines are present (\eg 1640\AA, 4686\AA), but
as soon as the gas metallicity goes up ((2)) also metal lines become
detectable, \eg\/ [OIII] 5007\AA\/ which is the strongest metal line in
the spectrum and  whose intensity is about 10\% of that of the H$\beta$
line for a 10$^{-3}$Z$_\odot$ metallicity. When the metal content in
the stars increases (case (3)), their effective temperatures are not
high enough to ionize He twice and, therefore, one finds weak, but
detectable, metal lines while HeII emission lines have disappeared. 
Finally, for high metallicities the emission line spectrum is dominated
by  metal lines, as customarily found in local Universe HII regions and
Planetary Nebulae.

More in detail, we note that both the luminosity and the equivalent
width of the  Ly-$\alpha$ line increase monotonically with decreasing 
metallicity (\cf Figures 4 and 5). In particular, we note that
equivalent widths in excess of 1,000\AA\/ are possible already for
objects with metallicity $\sim 10^{-3} Z_\odot$. This is particularly
interesting given that  Ly-$\alpha$ emitters with large EW have been
identified at z=5.6 (Rhoads \& Malhotra 2001).

   \begin{figure}
   \plotfiddle{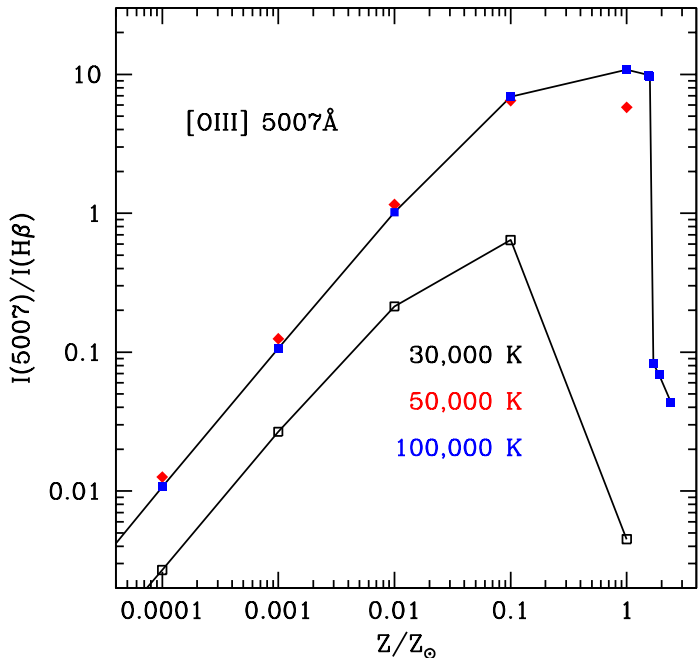}{7.99cm}{0}{50}{50}{-150}{-100}
   \caption{
   The ratio I([OIII]$\lambda 5007$)/I(H$\beta$) is plotted as a
function of metallicity for three different effective temperatures:
30,000 K (open squares and bottom line), 50,000 K (solid diamonds), and
100,000 K (solid squares and top line).}
   \end{figure} 

Even though the metal lines at low metallicities are weak, some of them
can be used as metallicity tracers. In Figure 6 the intensity ratio of 
 [OIII]$\lambda 5007$ to H$\beta$ is plotted for a range of
stellar temperatures and metallicities. It is immediately apparent that
for $Z<0.03Z_\odot$  this line ratio traces metallicity linearly.  Our
reference value $Z = 10^{-3}Z_\odot$ corresponds to an intensity ratio
I([OIII])/I(H$\beta$) $\simeq$ 0.1. The weak dependence on the
effective temperature makes sure that this ratio remains a good
indicator of metallicity not only for populations with a top-heavy IMF
but also in the more general case of sources with a wide range of
stellar masses. And indeed these predictions match quite well the
observational results found for blue dwarf galaxies (Izotov \& Thuan
1998, and references therein; Panagia
\etal\/2002).

Another difference between zero-metallicity and low-metallicity HII
regions is that the latter may contain dust, which could absorb part of
the stellar and nebular  radiation and re-emit it in the far infrared.
For example, {\it assuming} that dust formation processes work equally
well at all metallicities, and, therefore, that the dust content of a
region is proportional to its metallicity,  the dust optical depth in 
a 10$^6$ \msun\/ cloud with  $Z=10^{-3} Z_\odot$  is expected be
$\tau\sim0.03$ at the  Ly-$\alpha$ wavelength and $\tau\sim0.01$ in
the visual.  As a consequence, while only about 2\% of the stellar and
non-resonant nebular radiation is absorbed, as much as 30\% of the 
 Ly-$\alpha$ line is likely to be absorbed (\eg Panagia \& Ranieri
1973). Since the  Ly-$\alpha$ luminosity is about 40\% of the total
luminosity of the system, as much as 10-15\% ot the total energy will
be absorbed by dust and re-emitted in the far IR (Panagia \etal 2002).

   \begin{figure}
   \plotfiddle{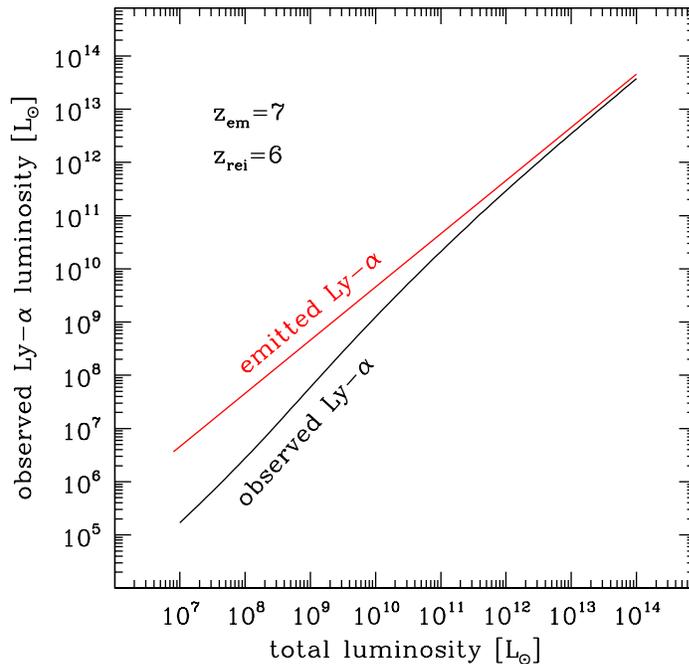}{7.99cm}{0}{50}{50}{-150}{-100}
   \caption{Transmitted  Ly-$\alpha$ intensity as a function of the object
luminosity. Brighter objects ionize their neighborhood and are
able to reduce the  Ly-$\alpha$ attenuation.}
   \end{figure} 

\section{Finding and characterizing  Primordial HII Regions}
\label{sect:future}

From an observational point of view one cannot measure a
zero-metallicity but one will usually be able to place an upper limit
to it. When is such an upper limit uniquely indicative of  a population
III object? In the previous sections we argue that a metallicity
Z$\simeq10^{-3}Z_\odot$ can be used as a dividing line between the pre
and post re-ionization phase of the Universe, and that essentially the
same value results from pollution by SNII explosions  in a primordial
cloud (see also Panagia \etal 2002). Thus, any object with a
metallicity higher than $\sim 10^{-3} Z_\odot$ is not a true first
generation object. Also, we  have seen which type of diagnostics we
have to characterize primordial and low-metallicity HII regions. 

Now next  question is: can we do it? do we have the means to detect
primordial sources and  verify that they are indeed pristine?

In this section we will focus on the capabilities of the NASA/ESA/CSA
Next Generation Space Telescope (NGST) that, thanks to its large
aperture  (6m class) and very low background (space-borne, passively
cooled to less than 50~K), is our best bet for success.

Before proceeding further, we have to consider the effects of
intergalactic HI absorption, \ie the well known Gunn-Peterson effect
(Gunn \& Peterson 1965): all radiation shortward of the Ly-$\alpha$
wavelength, as well as in the damping wing extending up to several
thousand \kms\/ longward of it,  is suppressed if IGM hydrogen is
mostly neutral. This could be a serious problem because the 
Ly-$\alpha$ line alone carries as much as 40-45\% of the total energy
and, by being concentrated in a narrow wavelength range, typically
$\Delta\lambda\simeq\lambda_0/1000\simeq1.2$\AA, is easily detectable
(\cf Figures 3 and 9). Fortunately, part of the same radiation that
ionizes the HII region can also leak out and ionize the surrounding
region of the IGM.  This may sound like a contrived {\it ad hoc}
hypothesis but an HII configuration that is density bounded in some
directions and ionization bounded in others is quite common in nature,
as well demonstrated by many galactic HII regions, such as M42 (the
Orion Nebula) or M8  (the Lagoon Nebula), etc.  It follows that for any
given leakage fraction, the higher is the ionizing photon flux of the
source, the bigger is the ionized sphere of IGM gas, and,  because of
cosmic expansion,  the damping wing of the  IGM HI trough shifts to
more amd more negative velocities, thus reducing the absorption on the
Lyman $\alpha$ emission profile  (Miralda-Escud\'e \& Rees 1998, Madau
\& Rees 2001, Panagia \etal 2002).   A comparison of the observed vs
emitted  Ly-$\alpha$ intensities is given in Figure 7.  The
transmitted  Ly-$\alpha$ flux depends on the total luminosity of the
source since this determines the radius of the resulting Str\"omgren
sphere. A  Ly-$\alpha$ luminosity of $\sim10^{10}$ L$_\odot$
corresponds to $\sim10^6$ M$_\odot$ in massive stars. In the following
we will consider a star cluster of this luminosity as our reference
case.

Now, if we  convolve the synthetic spectra derived above, including the
appropriate effect of neutral IGM absorption, with a broad band filter
response we obtain a spectral energy distribution that can be compared
directly to the NGST imaging sensitivity.  Figure 8 shows the
comparison of the flux from a $10^6$\msun\/ starburst at z=15,
convolved  with a bandwidth of    $\Delta\lambda/\lambda=5$,  with the
expected limiting flux that a 6m NGST can reach in an exposure of
$4\times10^5$s with a S/N=10. It is clear that NGST will be able to
easily detect such objects over the range 2--4 $\mu m$.

   \begin{figure}
    \plotfiddle{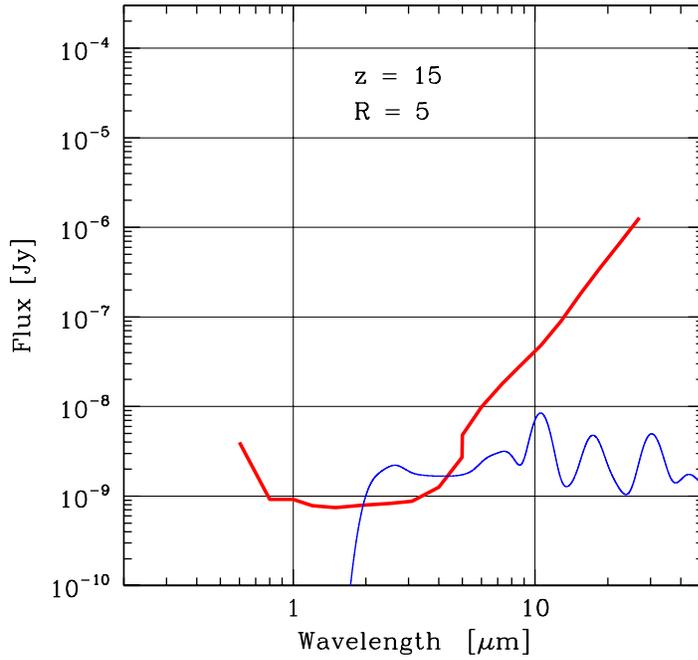}{7.5cm}{0}{50}{50}{-150}{-100}   
  \caption{
   Synthetic spectral energy distribution of a Z=$10^{-3} Z_\odot$
starburst object at z=15 containing $10^6$ M$_\odot$ in massive stars
(thin line) compared to the imaging limit of NGST at R=5 (thick line).
The NGST sensitivity refers to $4\times10^5$s exposures with S/N=10. }
   \end{figure} 

   \begin{figure}
   \plotfiddle{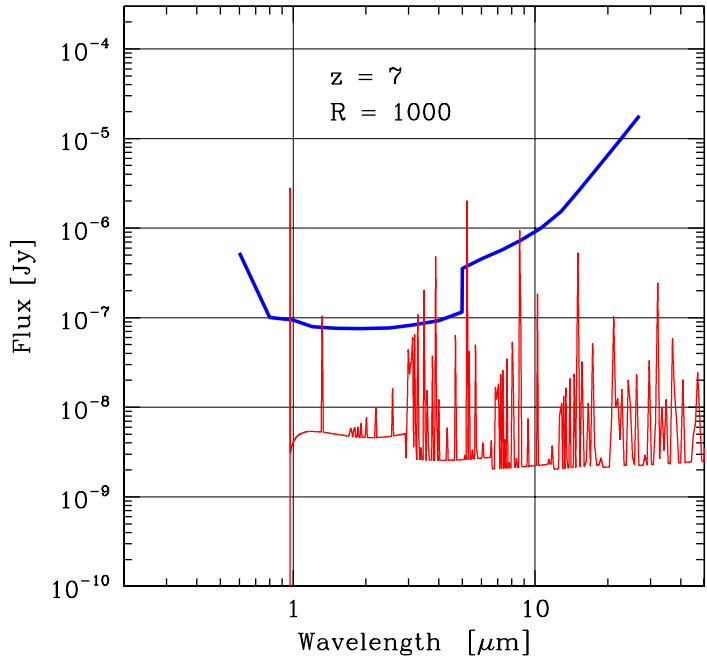}{7.5cm}{0}{50}{50}{-150}{-100}
   \caption{
   Synthetic spectrum of a Z=$10^{-3} Z_\odot$ starburst object at z=7
containing $10^6$ M$_\odot$ in massive stars (thin line) compared to the
spectroscopic limit of NGST at R=1000 (thick line). The NGST sensitivity
refers to $4\times10^5$ s exposures with S/N=10. }
   \end{figure} 

The synthetic spectra can also be compared to the NGST spectroscopic
sensitivity for $4\times10^5$s exposures (see Figure 9): it appears
that while the  Ly-$\alpha$ line can be detected up to redshifts as
high as 15 or 20, for our reference source only at relatively low
redshifts (z$\sim7$), can NGST detect other diagnostics lines lines
such as HeII 1640\AA, and Balmer lines. Determining metallicities is
then limited to either lower redshifts or to brighter sources.  Note
that a ``bright" source may be a more massive source but may also be a
{\it gravitationally lensed} object, of the type discovered by Ellis
\etal (2002) at redshift $\sim5.6$.

We can reverse the argument and ask ourselves what kind of sources can
NGST detect and characterize with spectroscopic observations.  Figure
10 displays, as a function of redshift,  the total luminosity of a
starburst that is  required  for a S/N=10 measurement of  a given line
with an exposure time of $4\times10^5$s. The loci for Ly-$\alpha$, HeII
1640\AA, H\/$\beta$, and [OIII] 5007\AA\/ are shown. It appears that
Ly-$\alpha$ is readily detectable up to z$\simeq$20 even for sources 10
fainter than our reference case, HeII 1640\AA\/ can also be detected up
to high redshifts {\it if} massive stars are indeed as hot as
predicted, whereas ``metallicity" information, \ie the intensity ratio
I([OIII])/I(H$\beta$), can be obtained at high redshifts only for
sources that either are 10--100 times more massive or are 10--100 times 
magnified by gravitational lensing.

   \begin{figure}
    \plotfiddle{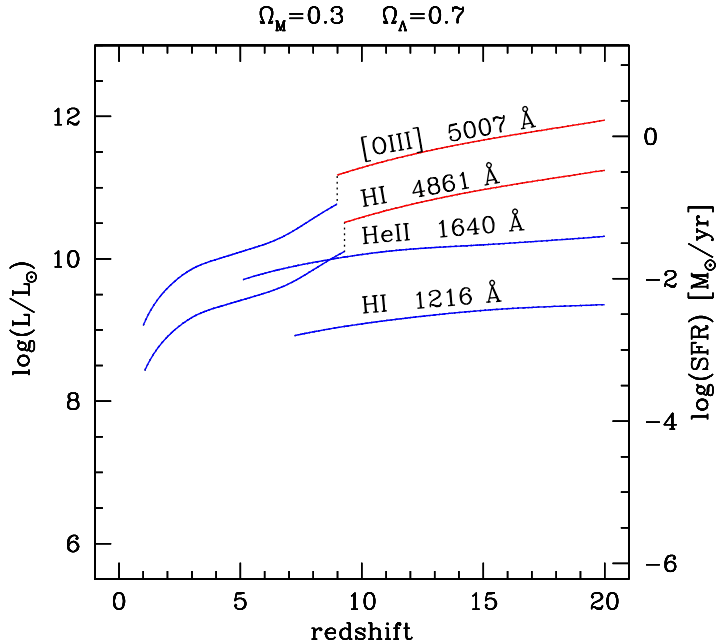}{7.99cm}{0}{50}{50}{-150}{-100}   
  \caption{Total source luminosities (left-hand scale) and massive
   star  formation rates (right-hand scale) required to detect selected
   emission lines radiated by Z=10$^{-3}$Z$_\odot$ HII regions using
   NGST R=1000 spectrometer with S/N=10 integrating for $4\times10^5$s.
   }
   \end{figure} 

\section{Conclusions}

It is possible to discern truly primordial populations from the next
generation of stars by measuring the metallicity of high-z star
forming objects. The very low background of NGST will enable it to
image first-light sources at very high redshifts, identifying them
through the Lyman break technique and/or spectroscopic detection of the 
 Ly-$\alpha$, and possibly HeII 1640\AA\/ emission. The relatively small
collecting area of a 6m NGST limits its capability in obtaining spectra
of z$\sim$10--15 first-light sources to either the bright end of their
luminosity function or to strongly lensed sources.

\end{document}